\documentclass[a4paper,11pt]{article}
\pdfoutput=1
\usepackage{jcappub}
\usepackage[T1]{fontenc}

\usepackage{tikz}
\usepackage{enumitem}
\usepackage{colortbl}
\usepackage{MnSymbol}

\title{Impact of half-wave plate systematics on the measurement\\ of CMB $B$-mode polarization}

\author[a]{Marta Monelli,}
\author[a,b,c]{Eiichiro Komatsu,}
\author[d]{Tommaso Ghigna,}
\author[c,e]{Tomotake Matsumura,}
\author[f,g]{Giampaolo Pisano,}
\author[c]{and Ryota Takaku.}

\affiliation[a]{
Max Planck Institute for Astrophysics, Karl-Schwarzschild-Str.\ 1, 85748 Garching, Germany
}
\affiliation[b]{
Ludwig-Maximilians-Universit\"at M\"unchen, Schellingstr.\ 4, 80799 M\"unchen, Germany
}
\affiliation[c]{
Kavli Institute for the Physics and Mathematics of the Universe (Kavli IPMU, WPI), UTIAS, The University of Tokyo, Chiba, 277-8583, Japan
}
\affiliation[d]{
International Center for Quantum-field Measurement Systems for Studies of the Universe and Particles (QUP, WPI), High Energy Accelerator Research Organization (KEK), Oho 1-1, Tsukuba, Ibaraki 305-0801, Japan
}
\affiliation[e]{
Center for Data-Driven Discovery, Kavli IPMU (WPI), UTIAS, The University of Tokyo,
Kashiwa, Chiba 277-8583, Japan
}
\affiliation[f]{Dipartimento di Fisica, Sapienza University of Rome, Rome, Italy}
\affiliation[g]{School of Physics and Astronomy, Cardiff University, CF24 3AA Cardiff, UK}

\emailAdd{monelli@mpa-garching.mpg.de}

\abstract{
Polarization of the cosmic microwave background (CMB) can help probe the fundamental physics behind cosmic inflation via the measurement of primordial $B$ modes.
As this requires exquisite control over instrumental systematics, some next-generation CMB experiments plan to use a rotating half-wave plate (HWP) as polarization modulator.
However, the HWP non-idealities, if not properly treated in the analysis, can result in additional systematics. 
In this paper, we present a simple, semi-analytical end-to-end model to propagate the HWP non-idealities through the macro-steps that make up any CMB experiment (observation of multi-frequency maps, foreground cleaning, and power spectra estimation) and compute the HWP-induced bias on the estimated tensor-to-scalar ratio, $r$. 
We find that the effective polarization efficiency of the HWP suppresses the polarization signal, leading to an underestimation of $r$. 
Laboratory measurements of the properties of the HWP can be used to calibrate this effect, but we show how gain calibration of the CMB temperature can also be used to partially mitigate it.
On the basis of our findings, we present a set of recommendations for the HWP design that can help maximize the benefits of gain calibration.
}

\newcommand{\changedONE}[2]{#2}
\newcommand{\changedTWO}[2]{#2}
\newcommand{\changedTHR}[2]{#2}
\newcommand{\changedFOU}[2]{#2}
\newcommand{\changedFIV}[2]{#2}
\newcommand{\changedSIX}[2]{#2}

\begin{document}

\maketitle
\flushbottom

\section{Introduction}
Observations of temperature anisotropies in the cosmic microwave background (CMB) have been crucial in shaping our current understanding of cosmology \cite{BOOMERANG:2005,Komatsu:2014ioa,Planck:2018cosmpar}. 
Valuable complementary information is encoded in polarization anisotropies, which have only been partially explored \cite{WMAPmaps,Planck:2018nkj,POLARBEAR:2019kzz,Polarbear:2020lii,ACT:2020gnv,SPT:2019nip,SPT-3G:2021eoc,BICEP:2021xfz,SPIDER:2021ncy}.
The main goal of the next generation of CMB experiments, involving both ground-based (Simons Observatory \cite{Ade_2019}, South Pole Observatory \cite{Moncelsi:2020ppj} and CMB Stage-4 \cite{Abazajian:2019eic}) and spaceborne (LiteBIRD \cite{LiteBIRD:2022cnt} and PICO \cite{NASAPICO:2019thw}) missions, is to probe the fundamental physics behind cosmic inflation \cite{PhysRevD.23.347,Sato:inflation,Linde:1981mu} by measuring primordial $B$-mode polarization~\cite{Kamionkowski:2015yta,Komatsu:2022nvu}. 

Inflation sources initial conditions for cosmological perturbations via primordial vacuum quantum fluctuations \cite{Mukhanov:1981xt,Guth:1982ec,Starobinsky:1982ee,Hawking:1982cz}. 
The relative amplitude of the resulting scalar and tensor perturbations is quantified in terms of the tensor-to-scalar ratio, $r$. 
Since tensor perturbations \cite{Grishchuk:1974ny,Starobinsky:1979ty} would leave a distinct $B$-mode signature on the CMB polarization \cite{Zaldarriaga:1996xe,Kamionkowski:1996ks,Seljak:1996gy,Kamionkowski:1996zd}, $r$ can be inferred from the angular power spectrum of the primordial $B$ modes.
To date, CMB observations have only placed upper bounds on $r$, the tightest being $r < 0.032$ (95\% CL) \cite{Tristram:2021tvh} (see also \cite{BICEP:2021xfz,Tristram:2020wbi,Campeti:2022vom}). 
Future surveys aim for unprecedentedly low overall uncertainties, which, depending on the true value of $r$, would lead to a detection or a tightening of the upper bounds, both of which would allow us to place strong constraints on inflationary models.

Such an ambitious goal can only be achieved through an exquisite control over systematics. 
To this end, some next-generation CMB experiments, including LiteBIRD, are planning to employ a rapidly spinning half-wave plate (HWP) as a polarization modulator, which can mitigate $1/f$ noise and reduce temperature-to-polarization leakage \cite{Johnson_2007,2010SPIE.7741E..1CR,ABS:2013dqh,Rahlin:2014rja,Misawa:2014hka,Hill:2016jhd,Takakura:2017ddx,2016SPIE.9914E..0JG,Bryan:2015qwa,2016RScI...87i4503E}.
However, any realistic HWP is characterized by non-idealities \cite{2010SPIE.7741E..2BB,pisano2014development,2017arXiv170602464A} that can induce additional systematics if not properly accounted for in the analysis \cite{Monelli:2022pru,Giardiello:2021uxq,Duivenvoorden:2020xzm,Matsumura:2014dda,Bao:2015eaa,Verges:2020xug,Patanchon:2023ptm}.

In this paper, we present a simple framework to propagate the HWP non-idealities through the three macro-steps that characterize any CMB experiment: observation of multi-frequency maps, foreground cleaning, and power spectra estimation. 
We exploit the simplicity of the harmonic internal linear combination (HILC) foreground cleaning method \cite{PhysRevD.68.123523} to keep the treatment semi-analytical. 
This choice, along with our working assumptions, makes the analysis computationally inexpensive\footnote{The \changedONE{}{main} analysis for this paper takes \changedONE{less than}{around} three minutes to run on a 32 GB RAM laptop computer.} and reflects our intention to develop an intuitive understanding of how the HWP affects the observed CMB.

The remainder of this paper is organized as follows. 
In section \ref{sec:math_r} we generalize the arguments presented in \cite{Monelli:2022pru} and provide a simple model for multi-frequency maps observed through a rapidly spinning HWP. 
We then introduce the HILC foreground cleaning method and present the procedure we will use to infer $r$. 
In section \ref{sec:analysis}, we discuss the specific choices we make to model sky, noise, and beams, and present the results of the analysis in two cases. 
First, we assume that the HWP is ideal and verify that the pipeline recovers the input CMB signal. 
Second, we consider LiteBIRD-like instrument specifics and assume realistic HWPs. 
We find that, for our choice of HWPs and $r_\text{true}=0.00461$ in input, the HWP non-idealities introduce an effective polarization efficiency that suppresses the polarization signal, resulting in \changedONE{$\hat{r} = 0.0043\pm 0.0005$}{$\hat{r} = (4.30^{+0.56}_{-0.53})\times 10^{-3}$.} 
We also show how including gain calibration of the CMB temperature in the map model can partially mitigate this effect. 
In section \ref{sec:discussion},  we derive a set of design recommendations that can help maximize the benefits of the gain calibration step. 
We also review the simplifying assumptions underlying the model and briefly discuss how they might be relaxed. 
Conclusions and perspectives are presented in section \ref{sec:conclusions_r}.

\section{Mathematical framework}\label{sec:math_r}
In this section we present a simple model for  multi-frequency maps observed through a rapidly spinning HWP. 
We also introduce the HILC foreground cleaning method and derive an explicit expression for the $B$-mode angular power spectrum of its solution, $C_{\ell,\textsc{hilc}}^{BB}$, given the modeled multi-frequency maps. 
Finally, we present the methodology we use to estimate the tensor-to-scalar ratio parameter, $r$, from $C_{\ell,\textsc{hilc}}^{BB}$.

\subsection{Modeling the observed maps}\label{sec:math_obsmap}
We describe linearly polarized radiation\footnote{The standard cosmological model predicts that no circular polarization is produced at the surface of last scattering. Even beyond standard cosmology, none of the models that have been proposed to source circular polarization (see, for instance, \cite{Cooray:2002nm,Alexander:2008fp,Bavarsad:2009hm,Sadegh:2017rnr,Inomata:2018vbu,Vahedi:2018abn,Alexander:2019sqb,Bartolo:2019eac,Lembo:2020ufn}) allows for a significant signal. We therefore consider only linear polarization.} by the Stokes $I$, $Q$ and $U$ parameters defined in right-handed coordinates with the $z$ axis taken in the direction of the observer's line of sight (telescope boresight), according to the ``CMB convention'' \cite{diSeregoAlighieri:2016lbr}. 
Given an incoming Stokes vector $\mathbf{S}\equiv(I,Q,U)$, the effect of a polarization-altering device on $\mathbf{S}$ can be described by a Mueller matrix $\mathcal{M}$, so that $\mathbf{S}'=\mathcal{M}\mathbf{S}$ \cite{bass2009handbook}. 
Assuming azimuthally symmetric and purely co-polarized beams, we can approximate the entire telescope's optical chain by means of a Mueller matrix acting on appropriately smoothed input Stokes parameters. 

This setup allows us to write the telescope response matrix\footnote{The response matrix, $A$, relates the sky maps to the time-ordered data, i.e.\ the collection of signals observed by all the instrument's detectors. $A$ encodes information about the telescope's pointings and the instrument specifics, such as the HWP Mueller matrix and the detectors' orientations.}, $A$, analytically, and to obtain simple expressions for both time-ordered data (TOD), $\mathbf{d}$, and binned maps, $\widehat{\mathbf{m}}$ \cite{Tegmark:1996qs}:
\begin{equation}
    \mathbf{d} = A \overline{\mathbf{m}} + \mathbf{n}\,,\qquad\qquad
    \widehat{\mathbf{m}} = \left(\widehat{A}^T\widehat{A}\right)^{-1} \widehat{A}^T\mathbf{d}\,,
\end{equation}
where $\overline{\mathbf{m}}$ denotes the pixelized $\{I,Q,U\}$ sky maps smoothed to the resolution of the instrument, $\mathbf{n}$ the noise contribution to the TOD, and $\widehat{A}$ the response matrix assumed by the map-maker. 

If the telescope's first optical element is a rapidly rotating HWP with Mueller matrix
\begin{equation}\label{eqn:HWPmueller}
    \mathcal{M}_\textsc{hwp} =
 \begin{pmatrix}
  m_\textsc{ii} & m_\textsc{iq} & m_\textsc{iu} \\
  m_\textsc{qi} & m_\textsc{qq} & m_\textsc{qu} \\
  m_\textsc{ui} & m_\textsc{uq} & m_\textsc{uu}
 \end{pmatrix},
\end{equation}
the maps reconstructed from the TOD of the $i$ channel's detectors by an ideal binning map-maker that assumes $\widehat{\mathcal{M}}_\textsc{hwp}=\text{diag}(1,1,-1)$ read\footnote{Eq.\ \eqref{eqn:widehatmi} follows from eq.\ (4.3) of \cite{Monelli:2022pru} by relaxing single frequency, CMB only, and no-noise assumptions.}
\begin{equation}\label{eqn:widehatmi}
    \widehat{\mathbf{m}}^i \!\simeq\! \sum_\lambda\! \int_{\nu^i_\text{min}}^{\nu^i_\text{max}}
    \! \frac{\text{d}\nu}{\Delta\nu^i}\, \begin{pmatrix}
    m_\textsc{ii}(\nu) & \phantom{-}0 & 0 \\
    0 & \!\phantom{-}[m_\textsc{qq}(\nu) - m_\textsc{uu}(\nu)]/2 & [m_\textsc{qu}(\nu) + m_\textsc{uq}(\nu)]/2 \\
    0 & \!-[m_\textsc{qu}(\nu) + m_\textsc{uq}(\nu)]/2 & [m_\textsc{qq}(\nu) - m_\textsc{uu}(\nu)]/2
    \end{pmatrix}\, \overline{\mathbf{m}}^{\,i}_\lambda(\nu) + \mathbf{n}^i,\!
\end{equation}
where the sum over $\lambda$ spans different sky components (CMB, dust, and synchrotron emission), the integral represents a top-hat bandpass with a bandwidth of $\Delta\nu^i\equiv \nu^i_\text{min} - \nu^i_\text{max}$, the superscript $i$ in $\overline{\mathbf{m}}^{\,i}_\lambda$ stresses that the input map is smoothed with the beam of the frequency channel $i$, and $\mathbf{n}^i$ denotes the noise maps.

Eq.\ \eqref{eqn:widehatmi} approximates the observed maps well when the cross-linking is good, that is, when each sky pixel is observed with a variety of scan angles. 
This condition is ensured by the rapid HWP rotation and the good LiteBIRD sky coverage, which guarantee that the scan angles are sampled uniformly enough for each pixel \cite{Monelli:2022pru}. 
As a consequence, our model neglects intensity-to-polarization leakage, the effects of which have been shown to be correctable \cite{Patanchon:2023ptm}.

If we also make the simplifying assumption that the spectral energy distribution (SED) of each component is uniform throughout the sky, we can rewrite each sky map as $\overline{\mathbf{m}}_\lambda(\nu)\equiv a_\lambda(\nu)\overline{\mathbf{m}}_\lambda(\nu_*)$, where $\nu_*$ is some reference frequency. 
This is equivalent to using the \textit{s0d0} option in the Python Sky Model (\texttt{PySM}) package~\cite{Thorne:2016ifb}, which has often been used in the literature for the study of systematics (e.g., \cite{LiteBIRD:2021hlz,Ghigna:2020wat}). 
The reason for this assumption is twofold. 
First, it is often useful to separate the effects of systematics from the complexity of the foreground emission. 
Second, as shown in \cite{LiteBIRD:2021hlz}, the study of systematics is strongly influenced by the specific \textit{class} of component separation methods, that is, whether it is a blind method, such as HILC \cite{PhysRevD.68.123523}, or a parametric method, such as \texttt{FGbuster} \cite{FGBuster}. 
In this paper, we use HILC and leave the study based on a parametric method for future work.

The factorization, $\overline{\mathbf{m}}_\lambda(\nu)= a_\lambda(\nu)\overline{\mathbf{m}}_\lambda(\nu_*)$, allows us to rewrite eq.\ \eqref{eqn:widehatmi} as
\begin{equation}\label{eqn:widehatmi'old}
    \widehat{\mathbf{m}}^i \simeq \sum_\lambda \begin{pmatrix}
    g_{\lambda}^{i} & 0 & 0 \\
    0 & \rho_{\lambda}^{i} & \eta_{\lambda}^{i} \\
    0 & -\eta_{\lambda}^{i} & \rho_{\lambda}^{i}
    \end{pmatrix}\, \overline{\mathbf{m}}^{\,i}_\lambda + \mathbf{n}^i\,,
\end{equation}
where we have dropped the $\nu_*$ dependence for the sake of simplicity and defined
\begin{subequations}\label{eqn:coeffgrhoeta}
\begin{align}
    g_{\lambda}^{i} &\equiv \int_{\nu^i_\text{min}}^{\nu^i_\text{max}}
    \! \frac{\text{d}\nu}{\Delta\nu^i} a_\lambda(\nu) m_{\textsc{ii}}(\nu)\,,\\
    \rho_{\lambda}^{i} &\equiv \frac{1}{2}\int_{\nu^i_\text{min}}^{\nu^i_\text{max}}
    \! \frac{\text{d}\nu}{\Delta\nu^i} a_\lambda(\nu) \left[m_{\textsc{qq}}(\nu)-m_{\textsc{uu}}(\nu)\right]\,,\\
    \eta_{\lambda}^{i} &\equiv \frac{1}{2}\int_{\nu^i_\text{min}}^{\nu^i_\text{max}}
    \! \frac{\text{d}\nu}{\Delta\nu^i} a_\lambda(\nu) \left[m_{\textsc{qu}}(\nu)+m_{\textsc{uq}}(\nu)\right]\,.
\end{align}
\end{subequations}
The coefficients in these equations have a clear physical interpretation: $g_{\lambda}^{i}$ is an effective gain for the temperature data, $\rho_{\lambda}^{i}$ and $\eta_{\lambda}^{i}$ are effective polarization gain (or polarization efficiency) and cross-polarization coupling, respectively, caused by the non-idealities of the HWP.

\paragraph{Including photometric calibration}
Photometric calibration is a crucial step in any CMB analysis pipeline that allows us to map the instrumental output to the incoming physical signal \cite{BeyondPlanck:2020bbb}. 
Here, we assume that the CMB temperature dipole \cite{Lineweaver:1996xa, piatDIPOLE} is used as a calibrator, as is commonly done in CMB experiments, and we neglect any imperfections in calibration. 
In other words, we assume to know $\tilde{g}^i = g^i_\text{CMB}$ exactly after calibration. 
The photometrically calibrated counterpart of eq.\ \eqref{eqn:widehatmi'old} reads
\begin{equation}\label{eqn:widehatmi'}
    \widehat{\mathbf{m}}^i \simeq \frac{1}{g^i_\text{CMB}}\left[\sum_\lambda \begin{pmatrix}
    g_{\lambda}^{i} & 0 & 0 \\
    0 & \rho_{\lambda}^{i} & \eta_{\lambda}^{i} \\
    0 & -\eta_{\lambda}^{i} & \rho_{\lambda}^{i}
    \end{pmatrix}\, \overline{\mathbf{m}}^{\,i}_\lambda + \mathbf{n}^i\right]\,.
\end{equation}
\paragraph{Spherical harmonics coefficients}
To apply the HILC method to the modeled maps, we expand eq.\ \eqref{eqn:widehatmi'} in spin-0 and spin-2 spherical harmonics and write the corresponding $B$-mode spherical harmonics coefficients as
\begin{equation}\label{eqn:alm_hat}
    \widehat{a}_{\ell m}^{B,i} = \frac{1}{g_\text{CMB}^i} \left[\sum_\lambda B^i_\ell\left( \rho_\lambda^i a^{B,i}_{\ell m} - \eta_\lambda^i a^{E,i}_{\ell m}\right) + n_{\ell m}^{B,i}\right]\,,
\end{equation}
where $a_{\ell m,\lambda}^{E}$ and $a_{\ell m,\lambda}^{B}$ are the $E$- and $B$-mode coefficients of the unsmoothed maps at some reference frequency $\nu_*$ (implicit here), and $B_\ell^i$ is the beam transfer function of the channel $i$. 

\subsection{Harmonic internal linear combination}\label{sec:HILC}
The internal linear combination (ILC) \cite{bennet_et_al} is a blind foreground cleaning method. 
It can be implemented in both map and multipole space, the latter case being referred to as HILC~\cite{PhysRevD.68.123523}. 
Given the spherical harmonics coefficients, $a_{\ell m}^{X,i}$ with $X=(T,E,B)$ and $i\in\{1,\dots,n_\text{chan}\}$, of the maps observed by each of the $n_\text{chan}$ frequency channels, the HILC solution is given by \cite{PhysRevD.68.123523}
\begin{equation}\label{eqn:sol_HILC}
    a^X_{\ell m, \textsc{hilc}} = \sum_{i=1}^{n_\text{chan}} w^{i}_{\ell}
    a^{X,i}_{\ell m}\,,\quad \text{with weights}\quad \mathbf{w}_\ell = \frac{\mathbb{C}_\ell^{-1} \mathbf{e}}{\mathbf{e}^T \mathbb{C}_\ell^{-1} \mathbf{e}}\,,
\end{equation}
where \changedFIV{}{$\mathbf{e}$ is a column vector with $n_\text{chan}$ elements all equal to one, and} $\mathbb{C}_\ell$ is the $n_\text{chan}\times n_\text{chan}$ covariance matrix of the observed maps: $\mathbb{C}_\ell^{ij} = \langle a_{\ell m}^{i*} a_{\ell m}^j\rangle$. 

By construction, the weights minimize the variance of the final map and add to unity, $\sum_iw_\ell^i=1$, preserving the frequency independence of the CMB black-body spectrum. 
However, the frequency dependence of $g_\mathrm{CMB}^i$, $\rho_\mathrm{CMB}^i$, and $\eta_\mathrm{CMB}^i$ can violate this sum rule. This is the main point we study in this paper.

\paragraph{Modeling the HILC solution}
To apply the HILC to the analytical predictions discussed in section \ref{sec:math_obsmap}, we could simply use eq.\ \eqref{eqn:alm_hat}; however, since different channels are characterized by different beams, it is preferable to perform the HILC on unsmoothed spherical harmonic coefficients, $a_{\ell m}^i \equiv \widehat{a}_{\ell m}^{B,i}/B_\ell^i$ and write the covariance matrix as
\begin{equation}\label{eqn:obs_cov}
    {\mathbb{C}}_{\ell}^{B,ij} = \frac{1}{g_\text{CMB}^ig_\text{CMB}^j} \left\{
    \sum_\lambda \left[\rho_\lambda^i\rho_\lambda^j C^{BB}_{\ell,\lambda} + \eta_\lambda^i\eta_\lambda^j C^{EE}_{\ell,\lambda}-\left(\rho_\lambda^i\eta_\lambda^j + \eta_\lambda^i \rho_\lambda^j\right) C_{\ell,\lambda}^{EB}\right] +\frac{\mathbb{N}_\ell^{BB,ij}
    }{B_\ell^iB_\ell^j}\right\}.
\end{equation}
We use eq.\ \eqref{eqn:obs_cov} to compute the HILC weights, $\mathbf{w}_\ell$, and the spherical harmonics coefficients of the HILC solution according to eq.\ \eqref{eqn:sol_HILC}. The corresponding angular power spectrum reads
\begin{equation}\label{eqn:sol_Cl_HILC}
    C_{\ell, \textsc{hilc}}^{BB}= \sum_{i,j=1}^{n_\text{chan}} \frac{w^{i}_{\ell} w^{j}_{\ell}}{g_\text{CMB}^i g_\text{CMB}^j} \left\{\sum_\lambda \Bigl[\rho_\lambda^i\rho_\lambda^j C^{BB}_{\ell,\lambda} + \eta_\lambda^i\eta_\lambda^j C^{EE}_{\ell,\lambda}-\Bigl(\rho_\lambda^i\eta_\lambda^j + \eta_\lambda^i \rho_\lambda^j\Bigr) C_{\ell,\lambda}^{EB}\Bigr] + \frac{\mathbb{N}_\ell^{BB,ij}}{B_\ell^iB_\ell^j}\right\}\,.
\end{equation}
This is the main equation from which we derive all of our results.

Even at this early stage, we can make some educated guesses about which terms will contribute the most to the final angular power spectrum.
By construction, the HILC tries to select the component $\lambda$ whose $\rho_\lambda^i$ and/or $\eta_\lambda^i$ are nearly constant across all frequency channels, i.e., a black-body spectrum. 
For example, if $m_\textsc{qq}(\nu)-m_\textsc{uu}(\nu)$ or $m_\textsc{qu}(\nu)+m_\textsc{uq}(\nu)$ depended on frequency as the inverse of the SED of the foreground emission, the foreground would leak into the HILC solution. 
However, the Mueller matrix elements of realistic HWPs do not exhibit such behavior. 
We therefore expect foreground-to-CMB leakage to be small in the final angular power spectrum.

Focusing on the CMB, eq.\ \eqref{eqn:sol_Cl_HILC} tells us that there are two potential contaminations: $E$-to-$B$ leakage, which can occur if the effective cross-polarization coupling, $\eta_\text{CMB}^i$, is nearly constant across the frequency channels, and suppression of the $B$ modes, which is instead driven by the effective polarization efficiency, $\rho_\text{CMB}^i$. 
The relative importance of these effects depends on the specific design choice of the HWP.

\subsection{Maximum likelihood estimate of the tensor-to-scalar ratio}\label{sec:MLE}
The modeled angular power spectrum is 
\begin{equation}
C^{BB}_{\ell}(r, A_\text{lens})= r C_\ell^\text{GW} + A_\text{lens} C_\ell^\text{lens} + N_\ell^{BB}\,, 
\end{equation}
where $C_\ell^\text{GW}$ is the primordial $B$-mode power spectrum with $r=1$~\cite{Seljak:1996gy,Kamionkowski:1996zd}, $C_\ell^\text{lens}$ is the lensed $B$-mode power spectrum~\cite{Zaldarriaga:1998ar}, $A_\text{lens}$ is its amplitude with $A_\text{lens}=1$ being the fiducial value, and $N_\ell^{BB}$ is the HILC solution for the total noise power spectrum [the last term in eq.~\eqref{eqn:sol_Cl_HILC}].

The probability density function (PDF) of the observed $B$-mode power spectrum for a given value of $r$ and $A_\text{lens}$, $P(C^{BB}_{l,\text{obs}}\,\vert\,r,A_\text{lens})$, is given by (e.g., \cite{Katayama:2011eh})
\begin{eqnarray}
\nonumber
    \log P(C^{BB}_{\ell,\text{obs}}\,\vert\, r,  A_\text{lens})  = &-&f_\text{sky} \frac{2\ell +1}{2} \left[\frac{C_{\ell,\text{obs}}^{BB}}{C_{\ell}^{BB}\!(r,  A_\text{lens})} + \log C_{\ell}^{BB}(r,  A_\text{lens}) -\frac{2\ell -1}{2\ell +1}\log C_{\ell,\text{obs}}^{BB}\right]\\
    &+&\mathrm{const.}\,,
\end{eqnarray}
where $f_\text{sky}$ is the sky fraction used to evaluate $C_{\ell,\text{obs}}^{BB}$. 
We use $f_\text{sky}=0.78$, for which our sky model is defined (see table \ref{tab:specs'} for details). 
Given the PDF, the likelihood function is
\begin{equation}\label{eqn:likelihood}
    L(r, A_\text{lens}) \propto \prod_{\ell=\ell_\text{min}}^{\ell_\text{max}}  P(C^{BB}_{\ell,\text{obs}}\,\vert\, r, A_\text{lens})\,.
\end{equation}
We use $\ell_\text{max} = 200$, which is the fiducial value for LiteBIRD \cite{LiteBIRD:2022cnt}. 
Using Bayes' theorem, the posterior PDF of $r$ with $A_\text{lens}$ marginalized over a flat prior is 
\begin{subequations}\label{eqn:likelihood_mp}
\begin{equation}\label{eqn:likelihood_m}
    L_\text{m}(r) \propto \int \text{d}A_\text{lens} \, L(r, A_\text{lens})\,.
\end{equation}
The frequentist profile likelihood is given instead by maximizing the bidimensional likelihood with respect to $A_\text{lens}$ for a set of values $\{r_0,\dots, r_n\}$ 
\begin{equation}\label{eqn:likelihood_p}
    L_\text{p}(r_i) \propto \max [L(r_i, A_\text{lens})]\,.
\end{equation}
\end{subequations}
Regardless of whether $L(r)\equiv L_\text{m}(r)$ or $L(r)\equiv L_\text{p}(r)$ is chosen, we define $\hat{r}$ as the maximum-likelihood estimate (MLE), i.e., the value of $r$ that maximizes $L(r)$. 
We compute the corresponding uncertainty as \cite{Katayama:2011eh}
\begin{equation}\label{eqn:sigma}
    \sigma_r^2 = \int_0^\infty \text{d}r\, L(r) r^2 - \left[\int_0^\infty \text{d}r\, L(r) r\right]^2\,,
\end{equation}
where $L(r)$ is normalized as $\int_0^\infty \text{d}r\, L(r)=1$. 
Eq.\ \eqref{eqn:sigma} defines the variance associated with a Gaussian random variable\changedSIX{}{, which is characterized by a likelihood that is symmetric with respect to its maximum}. \changedSIX{We use eq.\ (2.15) whenever we compute $\sigma_r$, but we have also compared it with asymmetric $68\%$ CL intervals. In our case, they are equal to the first significant digit.}{More generally, however, $L(r)$ may be asymmetric, and we estimate uncertainties as asymmetric 68\% CL intervals.}

\section{Analysis}\label{sec:analysis}
We apply the framework presented in section \ref{sec:math_r} to extract the bias on $r$ caused by a particular choice of HWP design. 
Given $\mathcal{M}_\textsc{hwp}$, our code\footnote{\href{https://github.com/martamonelli/HWP_end2end}{\texttt{github.com/martamonelli/HWP\_end2end}}.} performs the following steps:
\begin{enumerate}[topsep=2mm,parsep=1mm,itemsep=1mm]
    \item Compute the covariance matrix, $\mathbb{C}_\ell^{B,ij}$, as in eq.\ \eqref{eqn:obs_cov},
    \item Invert $\mathbb{C}_\ell^{B,ij}$ to obtain the HILC weights, $w_\ell^i$, as in eq.\ \eqref{eqn:sol_HILC},
    \item Use the $w_\ell^i$ to compute the $BB$ spectrum of the HILC solution, $C_{\ell,\textsc{hilc}}^{BB}$, as in eq.\ \eqref{eqn:sol_Cl_HILC},
    \item Compute the two-dimensional likelihood $L(r, A_\text{lens})$ from $C_{\ell,\textsc{hilc}}^{BB}$, according to eq.\ \eqref{eqn:likelihood},
    \item Obtain the one-dimensional posterior PDF, $L_\text{m}(r)$, by marginalizing over $A_\text{lens}$, and the profile likelihood, $L_\text{p}(r)$, by maximization,
    \item Return $\hat{r}$ and $\sigma_r$, defined as in eq.\ \eqref{eqn:sigma}, computed from $L_\text{m}(r)$ and $L_\text{p}(r)$.
\end{enumerate}
To validate our end-to-end model and code, we first perform the analysis for an ideal HWP and then move on to more realistic cases. 
However, before presenting our results, we review the additional assumptions that go into the explicit computation of the HILC covariance matrix $\mathbb{C}_\ell^{B}$, with the exception of the HWP choice.

\paragraph{CMB, dust and synchtrotron spectral responses}
For maps in thermodynamic units, the $a_\lambda(\nu)$ functions entering in eqs.\ \eqref{eqn:coeffgrhoeta} read (see appendix \ref{sec:app_a} for a complete derivation)
\begin{subequations}\label{eqn:alambda}
\begin{align}
    a_{\text{CMB}}(\nu) &= 1\,,\\
    a_{\text{dust}}(\nu) &= \left(\frac{\nu}{\nu_\filledstar}\right)^{\beta_\text{dust}} \frac{B_\nu(T_\text{dust})}{B_{\nu_\filledstar}(T_\text{dust})} \frac{\nu_\filledstar^2}{\nu^2}\frac{x_\filledstar^2e^{x_\filledstar}}{x^2e^{x}} \frac{(e^{x}-1)^2}{(e^{x_\filledstar}-1)^2}\,,\\
    a_{\text{sync}}(\nu) &= \left(\frac{\nu}{\nu_\smallstar}\right)^{\beta_\text{sync}} \frac{\nu_\smallstar^2}{\nu^2}\frac{x_\smallstar^2e^{x_\smallstar}}{x^2e^{x}} \frac{(e^{x}-1)^2}{(e^{x_\smallstar}-1)^2}\,,
\end{align}
\end{subequations}
where $B_\nu(T)$ denotes a black-body spectrum at temperature $T$, $x\equiv h\nu/(k_B T_0)$ and $T_0=2.725$ K is the average temperature of the CMB~\cite{Fixsen2009}. 
The values of the remaining parameters entering in eqs.\ \eqref{eqn:alambda} are specified in table \ref{tab:specs'}.
\begin{table}[t]
    \centering
    \begin{tabular}{p{0.40\textwidth}p{0.1\textwidth}}
        \arrayrulecolor{gray!70}
        \multicolumn{2}{l}{\vspace{-2mm}Spectral parameters \vspace{2.5mm}}\\
        \hline
        \vspace{-2mm} CMB temperature $T_0$ & \vspace{-2mm} $2.725$ K\\
        Dust temperature $T_\text{dust}$ & $19.6$ K\\
        Dust spectral index $\beta_\text{dust}$ & $1.55$\\
        Dust reference frequency $\nu_\filledstar$ & $353$ GHz\\
        Synchrotron spectral index $\beta_\text{sync}$ & $-3.1$\\
        Synchtrotron reference frequency $\nu_\smallstar$ & $30$ GHz
    \end{tabular}
    \hfill
    \begin{tabular}{p{0.19\textwidth}p{0.09\textwidth}p{0.07\textwidth}}
        \arrayrulecolor{gray!70}
        $C_\ell^{XX}$ parameters \vspace{.5mm} & \hfill$q\, [\mu$K$
^2]$\hfill \vspace{.5mm} & \hfill$\alpha$\hfill \vspace{.5mm}\\
        \hline
        \vspace{-2mm} Dust $EE$ & \vspace{-2mm} \,\hfill323\hfill\, & \vspace{-2mm} \,\hfill$-0.40$\hfill\,\\
        Dust $BB$ & \,\hfill119\hfill\, &  \,\hfill$-0.50$\hfill\,\\
        Synchrotron $EE$ & \,\hfill2.3\hfill\, &  \,\hfill$-0.84$\hfill\,\\
        Synchrotron $BB$ & \,\hfill0.8\hfill\, &  \,\hfill$-0.76$\hfill\,\\
        \textcolor{white}{B} \\
        \textcolor{white}{B}
    \end{tabular}
    \caption{Left panel: SED parameters entering in eqs.\ \eqref{eqn:alambda} for each component as reported in \cite{Planck:2018yye}. Right panel: The power-law parameters for the angular power spectra of synchrotron and thermal dust emission entering in eq.\ \eqref{eqn:powelaw} as reported in \cite{Planck:2018yye} for the \texttt{Commander} \cite{Eriksen:2007mx} analysis with $f_\text{sky} = 0.78$.}
    \label{tab:specs'}
\end{table}

\paragraph{CMB, dust and synchtrotron angular power spectra}
The CMB angular power spectrum is computed with \texttt{CAMB} \cite{2011ascl.soft02026L} assuming the best-fit 2018 \emph{Planck} values for the cosmological parameters \cite{Planck:2018cosmpar}, except for the tensor-to-scalar ratio, which is set to $r_\text{true}=0.00461$. 
This is the same fiducial value as assumed in \cite{LiteBIRD:2022cnt}, and corresponds to Starobinsky’s $R^2$ inflationary model \cite{Starobinsky:1980te} with the $e$-folding value of $N_* = 51$. 

As for the polarized foreground emission, we parameterize their angular power spectra as a power law \cite{Planck:2018yye}
\begin{equation}\label{eqn:powelaw}
 D_\ell \equiv \frac{\ell(\ell +1) C_\ell}{2\pi} = q\left(\frac{\ell}{80}\right)^\alpha\,.
\end{equation}
Specific values of the parameters are reported in table \ref{tab:specs'} for both dust and synchrotron. 
Note that we neglect any intrinsic $EB$ correlation in the input, which is inaccurate (polarized dust emission has been observed to have non-zero $TB$ correlation \cite{Planck:2018gnk,Weiland:2019uwg}, which implies the presence of a $EB$ correlation~\cite{Huffenberger:2019mjx,Clark:2021kze}, and cosmic birefringence \cite{Komatsu:2022nvu} would also result in a non-zero $EB$). 
When presenting our results in section \ref{sec:realistic}, we comment on this assumption and argue that allowing non-zero $EB$ in input would not dramatically affect the analysis.

\paragraph{Instrument specifics}
To simulate LiteBIRD's design, we consider an instrument that mounts three different telescopes at low (LFT), medium (MFT), and high frequency (HFT). 
The specific frequency ranges of each telescope and frequency channel are taken from \cite{LiteBIRD:2022cnt}.

\paragraph{Noise covariance matrix}
Using a rotating HWP as polarization modulator suppresses the polarized $1/f$ noise component \cite{Johnson_2007}. 
Being left with white noise only, we parameterize $N_\ell^{BB,i}$ as \cite{Katayama:2011eh}
\begin{equation}
    N_{\ell}^{BB,i} = \left[\frac{\pi}{10800}\frac{n_{p}^{i}}{\mu\text{K}\, \text{arcmin}}\right]^2\mu\text{K}^2\, \text{str}\,,
\end{equation}
where $n_{p}^{i}$ is the noise in Stokes parameters $Q$ or $U$ per pixel with solid angle $\Omega_\text{pix} = 1$ arcmin$^2$. 
The specific values assumed for each $n_p^i$ are taken from \cite{LiteBIRD:2022cnt}.

\paragraph{Beams}
Since we assume the beams to be Gaussian and perfectly co-polarized, the $B_\ell^i$ coefficients only depend on the beam's full width at half maximum (FWHM). 
Specific FWHM values for each channel are taken from \cite{LiteBIRD:2022cnt}.

\subsection{Validation: ideal HWP}\label{sec:null}
An ideal HWP is described by a frequency-independent Mueller matrix with elements
\begin{equation}
    \mathcal{M}_\text{ideal} = \text{diag}(1,1,-1)\,.
\end{equation} 
In this case, the coefficients $g^i_\lambda$ and $\rho^i_\lambda$ reduce to the average of the correspondent $a_\lambda(\nu)$ function over the band $i$ [eq.~\eqref{eqn:coeffgrhoeta}], which we will denote $a^i_\lambda$. 
The $\eta_\lambda^i$ coefficients go instead to zero. 
According to eq.\ \eqref{eqn:widehatmi'}, the multi-frequency maps reduce to 
\begin{equation}
   \widehat{\mathbf{m}}^i \simeq \overline{\mathbf{m}}^{\,i}_\text{CMB} + \frac{1}{a^i_\text{CMB}}\left[\sum_{\lambda\neq\text{CMB}} a^i_\lambda\, \overline{\mathbf{m}}^{\,i}_\lambda + \mathbf{n}^i\right]\,.
\end{equation}
While the CMB component is not affected by the presence of the ideal HWP, the foreground emission suffers from a color correction, and the noise term is rescaled channel-by-channel. 
In this simple situation, the HILC should perform well and recover the CMB signal plus some noise bias given by
\begin{equation}\label{eqn:sol_Cl_HILC_noiseonly}
    N_{\ell, \textsc{hilc}}^{BB}= \sum_{i=1}^{n_\text{chan}} \left(\frac{w^{i}_{\ell}}{a_\text{CMB}^i B^i_\ell}\right)^2 \mathbb{N}_{\ell}^{BB,ii}\,.
\end{equation}
We should therefore check that, for $\mathcal{M}_\text{ideal} = \text{diag}(1,1,-1)$, the HILC output is in good agreement with the input CMB angular power spectrum, once the noise bias is removed.

\begin{figure}
    \centering
    \includegraphics{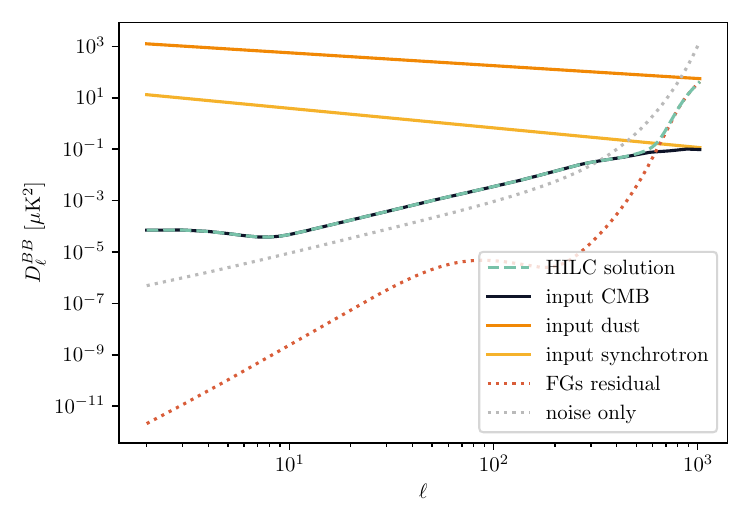}
    \vspace{-3mm}
    \caption{For an ideal HWP, the rescaled angular power spectrum, $D_\ell^{BB}$, of the HILC solution (dashed teal line) overlaps the input CMB spectrum (black solid line) for a wide range of multipoles. For large $\ell$, the two spectra begin to diverge as we approach the instrumental resolution. This can be seen by looking at the dotted gray line, representing the residual noise, which intersects the input spectrum at $\ell\sim 325$. For completeness, we also plot the input dust and synchrotron $D_\ell^{BB}$ (orange and yellow, respectively) and the foreground residual (red dotted line). The noise bias has been removed from both the HILC solution and the foreground residual spectra. The $w^i_\ell$ weights corresponding to the HILC solution are shown in figure \ref{fig:weights_IDEAL}. \label{fig:HILC_IDEAL}}
    \vspace{3mm}
    \begin{tikzpicture}
        \node at (10.4,0) {\includegraphics[scale=0.75, trim={0 .5cm 0 0}]{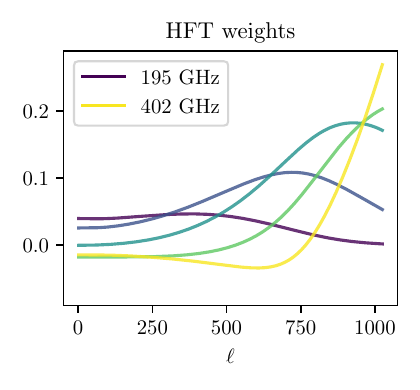}};
        \node at (5.2,0) {\includegraphics[scale=0.75, trim={0 .5cm 0 0}]{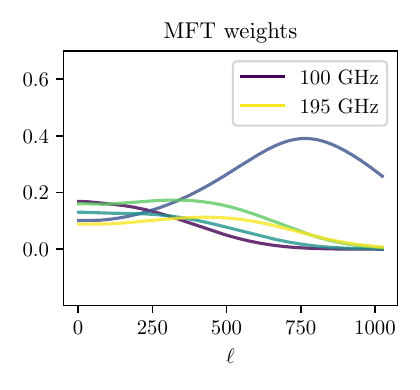}};
        \node at (0,0) {\includegraphics[scale=0.75, trim={.65cm .5cm 0 0}]{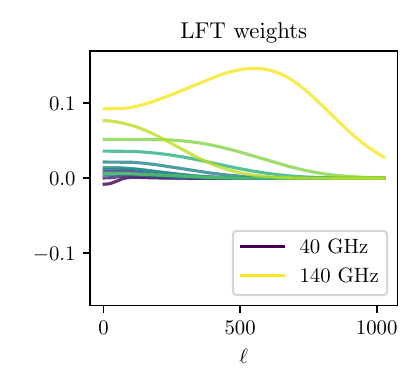}};
    \end{tikzpicture}
    \caption{HILC weights, $w_\ell^i$, for each of the three telescopes with an ideal HWP. \changedFOU{}{In each plot, different colored lines correspond to different frequency channels: from purple to yellow for lower to higher central frequencies (see \cite{LiteBIRD:2022cnt} for details on the channels' specifics).} The corresponding $BB$ angular power spectrum is shown in figure \ref{fig:HILC_IDEAL} (dashed teal line). \label{fig:weights_IDEAL}}
\end{figure}

In figure \ref{fig:HILC_IDEAL}, we show the angular $B$-mode power spectrum of the HILC solution, together with the input angular power spectra of CMB, dust, and synchrotron. 
For completeness, we also show the foreground residual and the noise bias. The noise bias has been removed from both the HILC solution and the foreground residual. 
The agreement between the HILC solution and the input CMB power spectrum is excellent up to $\ell\simeq 325$, roughly corresponding to LiteBIRD's beam resolution. 

In figure \ref{fig:weights_IDEAL} we show the HILC weights for the three telescopes. 
All MFT channels have positive weights, consistent with them being CMB channels. 
On the other hand, some of LFT and HFT channels (at very low and very high frequencies, respectively) have negative weights, resulting in foreground subtraction.

The code returns the MLE \changedONE{$\hat{r} = 0.0047 \pm 0.0005$}{$\hat{r} = (4.64^{+0.57}_{-0.54})\times 10^{-3}$}, 
which is compatible with the fiducial value of $r_\text{true}=0.00461$\changedTWO{}{, as the bias $\Delta r \equiv \hat{r}-r_\text{true}=0.03\times 10^{-3}$ is a small fraction of the uncertainty}. \changedSIX{}{Similarly, $A_\text{lens}$ is also unbiased: $\hat{A}_\text{lens}=1.00\pm 0.01$.} This is what we expect, given the good agreement between the debiased HILC solution and the input CMB shown in figure \ref{fig:HILC_IDEAL}.

\changedTHR{}{To test that the good agreement between the estimated $\hat{r}$ and $r_\text{true}$ is not just due to the specific value chosen for $r_\text{true}$, we repeat the analysis for a sample of the currently allowed values of $r_\text{true}$. The bias remains a small fraction of the error bar for all the values considered.} \changedONE{}{In particular, for $r_\text{true}=0$ the best fit is $\hat{r}=0$ with 68\% C.L. upper bound $0.00016$.
}

\subsection{More realistic HWPs}\label{sec:realistic}
For this analysis, we consider more realistic HWPs for each telescope. 
For LFT, we consider the Pancharatnam-type multi-layer sapphire symmetric stack design described in \cite{10.1117/1.JATIS.7.3.034005}, provided with an anti-reflection coating (ARC) as presented in \cite{LiteBIRD:2022gnn}. 
For the metal-mesh HWPs of MFT and HFT, we use the same input simulations and working assumptions as in \cite{Giardiello:2021uxq}. 

\begin{figure}
    \centering
    \includegraphics{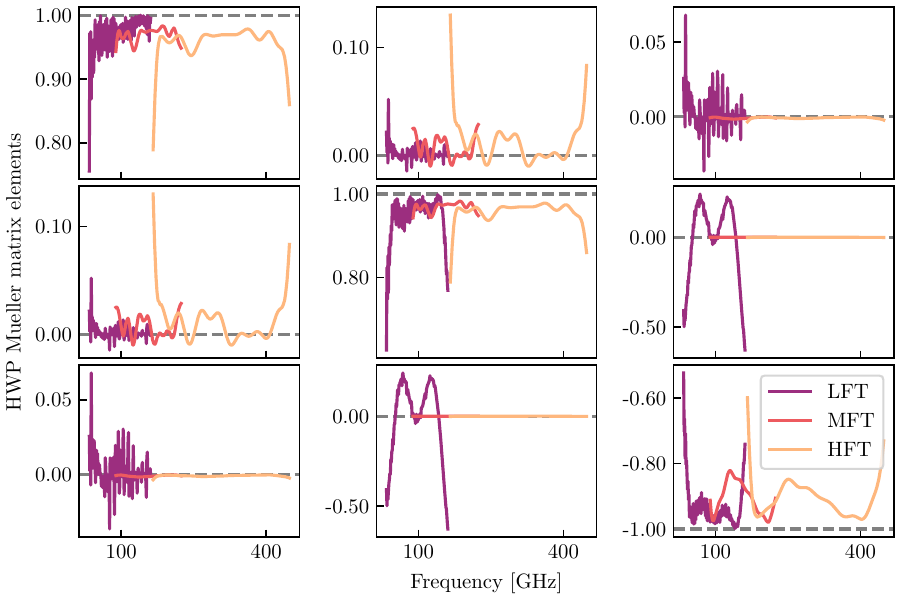}
    \caption{HWP Mueller matrix elements for LFT (purple), MFT (red) and HFT (orange) as function of frequency. For LFT, we consider a symmetric stack design \cite{10.1117/1.JATIS.7.3.034005} provided with ARC \cite{LiteBIRD:2022gnn}, compute its Mueller matrix elements, and rotate them of $55.02^\circ$, to express them in a reference frame with the $x$ axis parallel to the HWP optic axis. Instead, the Mueller matrix elements for MFT and HFT are obtained by following the same procedure and input simulations as done in \cite{Giardiello:2021uxq}, and rotating them of $0.29^\circ$. The dashed gray lines represent the ideal values of each element.}
    \label{fig:mueller}
\end{figure}

We manipulate each set of Mueller matrices by performing a rotation of the angle $\theta_\textsc{t}$ that minimizes the integral
\begin{equation}\label{eqn:integralrot}
    \int_\textsc{t} \text{d}\nu \, \left\{[m_{\textsc{qq}}(\nu) - m_\textsc{uu}(\nu)]\cos (4\theta_\textsc{t})  + [m_{\textsc{qu}}(\nu) + m_\textsc{uq}(\nu)]\sin(4\theta_\textsc{t})\right\}^2\,,
\end{equation}
over the entire frequency band of each telescope, specified by $\textsc{t}=\{\textsc{l},\textsc{m},\textsc{h}\}$.
This choice is ultimately motivated by the specific design we assume for LFT, since there is no unique way to determine the position of the HWP's optical axes for a symmetric stack. 
Rotating $\mathcal{M}_{\textsc{hwp},\textsc{l}}$ of $\theta_\textsc{l}$ then amounts to calibrate the HWP Mueller matrix and express it in a coordinate system aligned with the optical axes. 
Instead, the HWPs of MFT and HFT employ mesh-filter technology \cite{7460631}, for which optical axes can be more easily identified. 
However, for the sake of consistency, we choose to perform analogous rotations on the Mueller matrices of MFT and HFT metal-mesh HWPs. 
Rotation angles that minimize eq.\ \eqref{eqn:integralrot} are $55.02^\circ$ for LFT and $0.29^\circ$ for M-HFT. 
The rotated Mueller matrix elements of each HWP are shown as a function of frequency in figure \ref{fig:mueller}.

\begin{figure}
    \centering
    \includegraphics{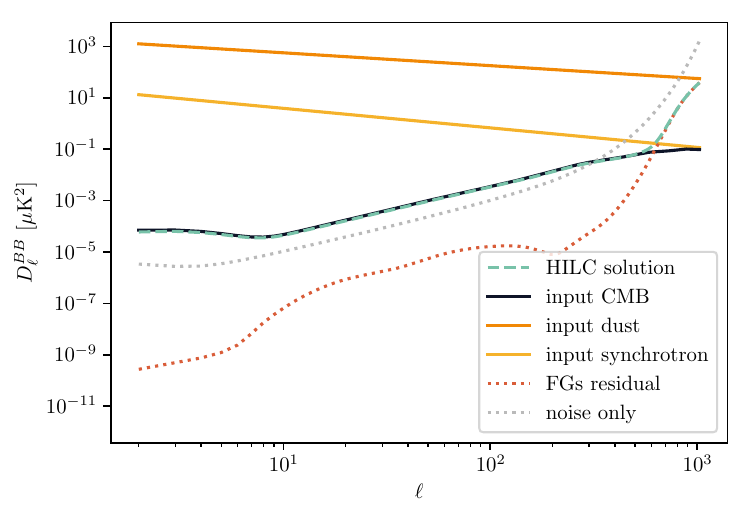}
    \vspace{-3mm}
    \caption{Same as figure~\ref{fig:HILC_IDEAL} but for the realistic HWP discussed in section \ref{sec:realistic} (dashed teal line). Compared to the ideal HWP case shown in figure \ref{fig:HILC_IDEAL}, the non-ideal HILC solution slightly differs from the input CMB at low multipoles. For comparison, we also show the residual noise bias (dotted gray line) and the foreground residual (red dotted line). They both show more features than their counterparts in figure \ref{fig:HILC_IDEAL}. The $w^i_\ell$ weights corresponding to the HILC solution are shown in figure \ref{fig:weights}. \label{fig:HILC}}
\end{figure}

Given the elements of the Mueller matrix, we compute the coefficients $\rho_\lambda^i$ and $\eta_\lambda^i$ according to eq.\ \eqref{eqn:coeffgrhoeta} and repeat all the steps outlined at the beginning of section \ref{sec:analysis}. 
The HILC solution, $D_{\ell,\textsc{hilc}}^{BB}$, is shown in figure \ref{fig:HILC}. 
Although the foreground residual (red dotted line) shows more features than in the ideal case of figure \ref{fig:HILC_IDEAL}, its contribution to $D_{\ell,\textsc{hilc}}^{BB}$ is still subdominant. 
This confirms our intuition that reasonably optimized HWPs do not cause strong foreground leakage in the HILC solution [see the discussion below eq.\ \eqref{eqn:sol_Cl_HILC}]. 
Note that, given the negligible foreground leakage, taking $C_{\ell,\text{dust}}^{EB} = C_{\ell,\text{synch}}^{EB} = 0$ in input is not such a strong assumption. 
Even if we allowed non-zero $EB$ correlations, they would not contribute significantly to the HILC solution.

\begin{figure}
    \centering
    \begin{tikzpicture}
        \node at (10.4,0) {\includegraphics[scale=0.75, trim={0 .5cm 0 0}]{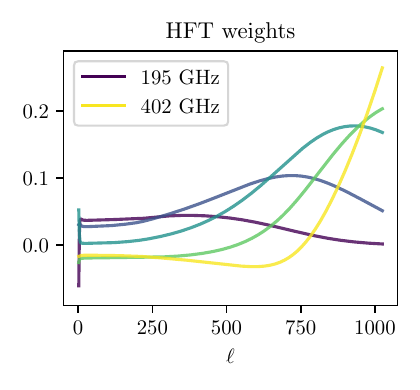}};
        \node at (5.2,0) {\includegraphics[scale=0.75, trim={0 .5cm 0 0}]{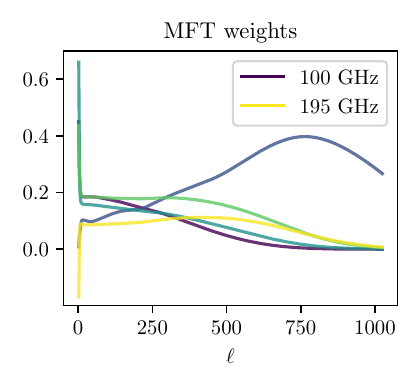}};
        \node at (0,0) {\includegraphics[scale=0.75, trim={.65cm .5cm 0 0}]{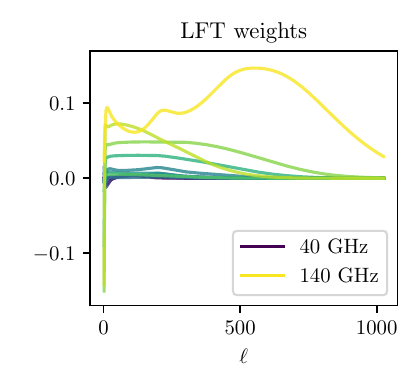}};
    \end{tikzpicture}
    \caption{Same as figure~\ref{fig:weights_IDEAL} but for the Mueller matrix elements given in figure \ref{fig:mueller}. The corresponding $BB$ angular power spectrum is shown in figure \ref{fig:HILC} (dashed teal line). \label{fig:weights}}
\end{figure}

In figure \ref{fig:weights} we also show the HILC weights for the three telescopes. 
The weights look qualitatively similar to their ideal counterparts shown in figure \ref{fig:weights_IDEAL}.

\begin{figure}
    \centering
    \includegraphics{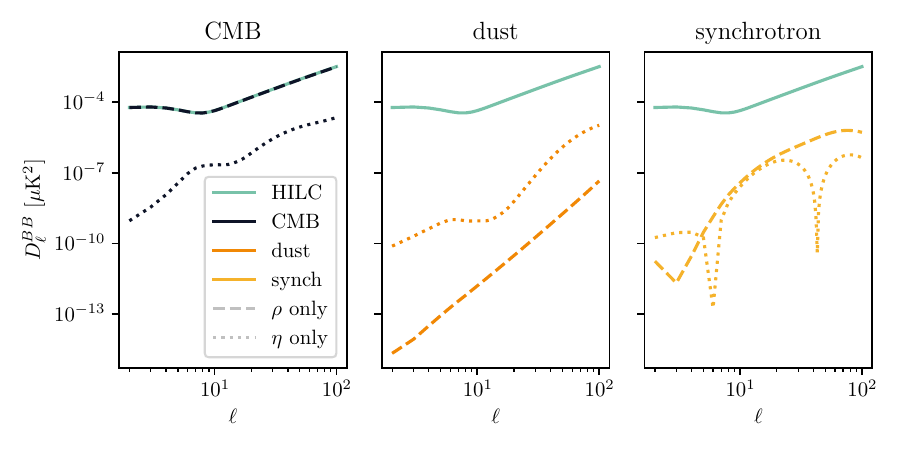}
    \caption{Different contributions to the $B$-mode power spectrum of the HILC solution (teal solid line). We focus on a different component (CMB, dust, and synchrotron) in each of the panels. The effective polarization efficiency and cross-polarization coupling components are shown in dashed and dotted, respectively. The largest contribution comes from the polarization efficiency component of the CMB.}
    \label{fig:HILC_components}
\end{figure}

To give more precise considerations, figure \ref{fig:HILC_components} shows the power spectra on large angular scales in more detail. 
We show the two independent terms that contribute to $D_{\ell,\textsc{hilc}}^{BB}$ component-by-component: $\rho$-only (polarization efficiency) and $\eta$-only (cross-polarization coupling). 
These were obtained using the full covariance matrix $\mathbb{C}_\ell$ given in eq.\ \eqref{eqn:obs_cov} to compute the HILC weights, while neglecting some of the terms entering in eq.\ \eqref{eqn:sol_Cl_HILC}. 
For instance, the $\rho$-only dust contribution reads
\begin{equation}
    C_{\ell, \textsc{hilc}}^{BB,\text{dust},\rho}= \sum_{i,j=1}^{n_\text{chan}} \frac{w^{i}_{\ell} w^{j}_{\ell}}{g_\text{CMB}^i g_\text{CMB}^j}     \rho_\text{dust}^i\rho_\text{dust}^j C^{BB}_{\ell,\text{dust}}\,.
\end{equation}
Intuitively, it makes sense for the effective polarization efficiency component to dominate in the CMB contribution.
While $\eta_\text{CMB}^i$ can be both positive and negative, all $\rho_\text{CMB}^i$ are constrained to be smaller than 1. 
This means that, while the average $\langle \eta_\text{CMB}^i \rangle$ across all frequency channels can be close to zero, $\langle \rho_\text{CMB}^i \rangle$ cannot be arbitrarily close to 1. 
The HILC, which looks for the solution that minimizes the variance, may then be able to get rid of all cross-polarization coupling, while it cannot undo the average suppression due to the polarization efficiency. 
As a consequence of the smallness of the cross-polarization coupling component relative to the polarization efficiency, we argue that relaxing the $C_{\ell,\text{CMB}}^{EB} = 0$ assumption for the input spectra would not significantly change our results.

Interestingly, the HILC solution approximately satisfies 
\begin{equation}
    \widehat{C}_{\ell,\textsc{hilc}}^{BB} \simeq \frac{1}{n_\text{chan}} \sum_{i=1}^{n_\text{chan}} \left[\frac{\rho_\text{CMB}^i}{g_\text{CMB}^i}\right]^2 \! \cdot \, C_{\ell,\text{CMB}}^{BB}\,,
\end{equation}
with $10^{-5}$ relative tolerance and $10^{-8}$ absolute tolerance for a wide range of multipoles, $25 \le \ell \le 372$. 
The upper limit has a simple interpretation: it roughly corresponds to the instrumental resolution.

\paragraph{Bias on the tensor-to-scalar ratio}
We finally employ the methodology introduced in section \ref{sec:MLE} to propagate the small discrepancy between the input CMB and the HILC solution shown in figure \ref{fig:HILC} into a bias on $r$. 
We compare the marginalized posterior PDF, $L_\text{m}(r)$, with the profile likelihood, $L_\text{p}(r)$ [as defined in eqs.\ \eqref{eqn:likelihood_m} and \eqref{eqn:likelihood_p}, respectively], and find that they are identical up to relative discrepancies of $\lesssim 10^{-3}$.

\begin{figure}
  \centering
  \includegraphics{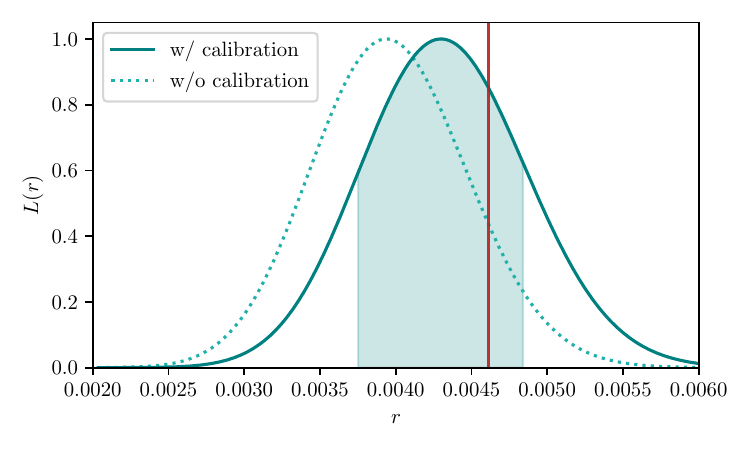}
  \caption{Normalized profile likelihood, $L(r) = L_\text{p}(r)$, obtained from the HILC solution, $\widehat{C}_{\ell,\textsc{hilc}}^{BB}$, given the HWP specifics presented in section \ref{sec:realistic} (teal solid line). The likelihood has a maximum at $\hat{r} = 0.0043\changedONE{}{0}$. The shaded region \changedONE{}{identifies the $68\%$ CL interval, and} goes from \changedONE{$\hat{r} - \sigma_r$}{$\hat{r} - 0.00053$} to \changedONE{$\hat{r} + \sigma_r$}{$\hat{r} + 0.00056$} \changedONE{where $\sigma_r = 0.0005$}{}. The solid red line represents the input tensor-to-scalar ratio parameter, $r_\text{true}=0.00461$. The dotted light teal line shows the normalized profile likelihood obtained from the HILC solution when the gain calibration for the CMB temperature is not included.}
  \label{fig:likelihoods}
\end{figure}

We show $L(r) = L_\text{p}(r)$ in figure \ref{fig:likelihoods} (teal solid line), together with a red vertical line corresponding to the input value, $r_\text{true}=0.00461$. 
The MLE is \changedONE{$\hat{r}=0.00430\pm 0.0005$}{$\hat{r}=(4.30^{+0.56}_{-0.53})\times 10^{-3}$}. 
\changedTWO{This bias}{The bias, $\Delta r = -0.31\times 10^{-3}$, is comparable to the uncertainty. We find that this bias} is caused by the HWP polarization efficiency being lower than one. The $B$-mode signal is suppressed and $r$ is underestimated. \changedSIX{}{Note that the suppression due to the HWP polarization efficiency also affects the observed lensing amplitude: $\hat{A}_\text{lens}=0.9548^{+0.0093}_{-0.0096}$.}

\changedONE{}{We also find non-detectable bias in the $r_\text{true}=0$ case: the best fit is $\hat{r}=0$ with 68\% C.L. upper bound $0.00017$, similarly to the ideal HWP case (see Section \ref{sec:null}).}

\paragraph{The weight of gain calibration}
The inclusion of the gain calibration for the CMB temperature in the modeling of multi-frequency maps may seem inconsequential, but it has strong implications. 
We repeat the analysis of section \ref{sec:realistic}, except that we now skip the gain calibration, i.e., we model the $\widehat{\mathbf{m}}^i$ as in eq.\ \eqref{eqn:widehatmi'old} instead of eq.\ \eqref{eqn:widehatmi'}. 
The corresponding spherical harmonic coefficients read
\begin{equation}\label{eqn:alm_hat_NOgain}
    \widehat{a}_{\ell m, \text{w/o}}^{B,i} = \sum_\lambda B^i_\ell\left( \rho_\lambda^i a^{B,i}_{\ell m} - \eta_\lambda^i a^{E,i}_{\ell m}\right) + n_{\ell m}^{B,i}\,,
\end{equation}
where the $\text{w/o}$ subscript stresses that we are not calibrating the maps. 
By retracing the same steps as presented in section \ref{sec:HILC}, we end up with an expression for the $BB$ angular power spectrum of the HILC solution that reads
\begin{equation}\label{eqn:sol_Cl_HILC_NOgain}
    C_{\ell, \textsc{hilc}}^{BB}= \sum_{i,j=1}^{n_\text{chan}} w^{i}_{\ell,\text{w/o}} w^{j}_{\ell,\text{w/o}} \left\{\sum_\lambda \Bigl[\rho_\lambda^i\rho_\lambda^j C^{BB}_{\ell,\lambda} + \eta_\lambda^i\eta_\lambda^j C^{EE}_{\ell,\lambda}-\Bigl(\rho_\lambda^i\eta_\lambda^j + \eta_\lambda^i \rho_\lambda^j\Bigr) C_{\ell,\lambda}^{EB}\Bigr] + \frac{\mathbb{N}_\ell^{BB,ij}}{B_\ell^iB_\ell^j}\right\}\,,
\end{equation}
where the $w^i_{\ell,\text{w/o}}$ are the HILC weights corresponding to the spherical harmonic coefficients of eq.\ \eqref{eqn:alm_hat_NOgain}. 
The corresponding normalized profile likelihood is shown in figure \ref{fig:likelihoods} (dotted light teal line). 
We now find a much lower MLE of the tensor-to-scalar ratio, \changedONE{$\hat{r}= (0.0039 \pm 0.0005$}{$\hat{r}= (3.94^{+0.52}_{-0.50})\times 10^{-3}$}, which is incompatible with $r_\text{true}$\changedTWO{}{, as the bias $\Delta r=-0.67\times 10^{-3}$ is larger than the uncertainty}. \changedSIX{}{Similarly, the bias on the lensing amplitude is also stronger than the case when photometric calibration is included: $\hat{A}_\text{lens}=0.913\pm{0.009}$.}

\section{Discussion}\label{sec:discussion}
Clearly, gain calibration can partially mitigate the suppression of primordial $B$ modes caused by the HWP. 
Of course, one can characterize the non-idealities in laboratory measurements and correct for them in the data. 
However, if HWPs are properly designed, gain calibration for the CMB temperature allows us to mitigate the effects of non-idealities on polarization \textit{in-flight} for space missions. 
The ability to perform in-flight calibration is always valuable. 

To this end, we derive some realistic recommendations that can help maximize its benefits. 
In section \ref{sec:assumptions}, we also discuss the assumptions underlying our end-to-end model and comment on the possibility of relaxing some of them.

\subsection{HWP design recommendations}\label{sec:HWPdesignREC}

We express the relevant combinations of Mueller matrix elements in terms of a set of 7 independent values that uniquely determine the components of $\mathcal{M}_\textsc{hwp}$: the HWP Jones parameters, $h_{1,2}$, $\beta$, $\zeta_{1,2}$ and $\xi_{1,2}$ (see appendix \ref{sec:muellerjones} for their definitions). 
The loss parameters $h_{1,2}$ describe the deviation from the unitary transmission of $E_{x,y}$; $\beta$ parametrizes the deviation from $\pi$ of the phase shift between $E_x$ and $E_y$; $\zeta_{1,2}$ and $\xi_{1,2}$ describe the amplitude and phase of the cross-polarization coupling. 
We write $g(\nu)\equiv m_\textsc{ii}(\nu)$, $\rho(\nu)\equiv [m_\textsc{qq}(\nu)-m_\textsc{uu}(\nu)]/2$, and $\eta(\nu)\equiv [m_\textsc{qu}(\nu)+m_\textsc{uq}(\nu)]/2$ as \cite{Giardiello:2021uxq}
\begin{subequations}\label{eqn:jones}
\begin{align}
    g &= \frac{1}{2}\left[(1+h_1)^2 + (1+h_2)^2 + \zeta_1^2 + \zeta_2^2\right]\,,\\
    \rho &= \frac{1}{2}\left\{\frac{1}{2}\left[(1+h_1)^2 + (1+h_2)^2 - \zeta_1^2 - \zeta_2^2\right] + (1+h_1)(1+h_2)\cos\beta - \zeta_1\zeta_2 \cos(\chi_1-\chi_2)\right\}\,, \\
    \eta &= \frac{1}{2}\left\{(1+h_1)(\zeta_1\cos\chi_1 +\zeta_2\cos\chi_2) + (1+h_2)\left[\zeta_2\cos(\beta-\chi_2) +\zeta_1\cos(\beta-\chi_1)\right] \right\}\,,
\end{align}
\end{subequations}
where any dependence on $\nu$ is kept implicit for the sake of compactness. 
Designing a perfectly ideal HWP with identically vanishing Jones parameters is technically impossible. 
However, some parameters are easier to minimize than others. 

For example, $\zeta_{1,2}(\nu)\sim 10^{-2}$ can be achieved for both metal-mesh and multi-layer HWPs. 
If that is the case, the Taylor expansion of the above expressions for small $\zeta_{1,2}(\nu)$ yields, up to first order,
\begin{subequations}\label{eqn:jones_zeta}
\begin{align}
    g &= \frac{1}{2}\left[(1+h_1)^2 + (1+h_2)^2\right] + \mathcal{O}(10^{-4})\,,\\
    \rho &= \frac{1}{2}\left\{\frac{1}{2}\left[(1+h_1)^2 + (1+h_2)^2\right] + (1+h_1)(1+h_2)\cos\beta\right\} + \mathcal{O}(10^{-4})\,, \\
    \eta &= \frac{1}{2}\left\{(1+h_1)(\zeta_1\cos\chi_1 +\zeta_2\cos\chi_2) + (1+h_2)\left[\zeta_2\cos(\beta-\chi_2) +\zeta_1\cos(\beta-\chi_1)\right] \right\}\,.
\end{align}
\end{subequations}
We can further simplify these expressions by requiring $h_{1,2}\sim 10^{-2}$, which implies $\rho(\nu)=g(\nu)\cos^2[\beta(\nu)/2]$ up to relative corrections of $\mathcal{O}(10^{-4})$. 
Alternatively, by keeping $h_{1,2}$ free while requiring $\vert h_1 - h_2\vert$ to be small, we ensure that $\rho(\nu)=g(\nu)\cos^2[\beta(\nu)/2]$ still holds up to relative corrections of $\mathcal{O}(\vert h_1 - h_2\vert)$. 
On the other hand, we cannot require $\beta(\nu)$ to be arbitrarily small due to the limitation of current technology. 
Keeping $\beta(\nu)$ free, we have
\begin{subequations}
\begin{align}
    g_\text{CMB}^{i} &\simeq \int_{\nu^i_\text{min}}^{\nu^i_\text{max}}
    \! \frac{\text{d}\nu}{\Delta\nu^i} \left[1+h_1(\nu)+h_2(\nu)\right]\,,\\
    \rho_\text{CMB}^{i} &\simeq \int_{\nu^i_\text{min}}^{\nu^i_\text{max}}
    \! \frac{\text{d}\nu}{\Delta\nu^i} \left[1+h_1(\nu)+h_2(\nu)\right]\cos^2[\beta(\nu)/2]\,.
\end{align}
\end{subequations}
If at least one of $h_{1}(\nu) + h_{2}(\nu)$ and $\cos^2[\beta(\nu)/2]=[1+\cos\beta(\nu)]/2$ is slowly varying within the band, we find that $\rho_\text{CMB}^i\simeq A^i\, g_\text{CMB}^i$, where $A^i$ is an appropriate factor that depends on $\beta$. 
Then, if we know $A^i$ with good precision, its effect can be undone by multiplying each multi-frequency polarization map by $1/A^i$. 
In this way, the gain calibration for the CMB temperature can partially mitigate the impact of the HWP polarization efficiency. 

Regarding cross-polarization coupling, we argue that there are two strategies to keep its effects under control. 
First, we could simply require $\eta(\nu)\lesssim 10^{-3}$ so that the $E\to B$ leakage is negligible. 
However, this might be technically challenging. 
Another strategy is to exploit the fact that the HILC weights minimize the variance. 
Even if $\eta(\nu)$ is not vanishing small, as long as the $\eta_\text{CMB}^i$ fluctuate around zero, the HILC should be able to mitigate their effect.

\paragraph{HWP angle miscalibration}
An imperfect calibration of the HWP angle can dramatically affect the considerations we have presented so far. 
If an HWP with $g^i_\text{CMB}\simeq \rho^i_\text{CMB}$ and $\langle\eta^i_\text{CMB}\rangle\simeq 0$, is rotated by some angle $\theta$, its effective gain, polarization efficiency, and cross-polarization coupling are transformed as
\begin{equation}
    g'=g\,,\qquad \rho'=\rho\cos4\theta - \eta\sin4\theta\,,\qquad \eta' = \eta\cos 4 \theta + \rho\sin 4 \theta\,.
\end{equation}
On the one hand, this causes the cross-polarization coupling coefficients to fluctuate around some non-zero value, making it impossible for the HILC to filter them out. 
On the other hand, the polarization efficiency and gain coefficients might strongly deviate from each other, reducing the benefits of gain calibration.

Therefore, a good calibration of the HWP position angle, $\theta$, is crucial to ensure the validity of our considerations and recommendations. 
Derotating the polarization maps by $\theta$ prior to the foreground cleaning step, as suggested in \cite{LiteBIRD:2021hlz}, would allow us to account for potential differences in the miscalibration angles of the HWPs.

\subsection{Reviewing the underlying assumptions}\label{sec:assumptions}
We derived the model for multi-frequency maps and their spherical harmonics coefficients [eqs.\ \eqref{eqn:widehatmi'} and \eqref{eqn:alm_hat}, respectively] under several assumptions. We list them in order of appearance:
\begin{enumerate}[topsep=2mm,parsep=1mm,itemsep=1mm]
    \item We assumed axisymmetric and perfectly co-polarized beams, \label{it:1}
    \item We assumed the maps to be obtained from an ideal bin averaging map-maker, \label{it:2}
    \item We considered a top-hat bandpass, \label{it:3}
    \item We assumed the SED of each component to be uniform throughout the sky, \label{it:4}
    \item We assumed a perfect gain calibration for the CMB temperature. \label{it:5}
\end{enumerate}
Assumptions \ref{it:1} and \ref{it:2} cannot be relaxed while maintaining the semi-analytical treatment, since more complex beams and more refined map-makers can only be included in numerical simulations. 
On the other hand, assumptions \ref{it:3} and \ref{it:5} can be straightforwardly relaxed within our simple analytical model (given our focus on the HWP non-idealities, however, we chose not to play around with the bandpass shape or imperfect temperature gain calibration). 

Assumption \ref{it:4} can also be relaxed easily, but allowed us to analytically model the foreground cleaning step. 
Indeed, as soon as the SED of the foreground emission becomes anisotropic, the simple implementation of the HILC presented in section \ref{sec:HILC} is no longer able to recover the CMB signal accurately, and more elaborate methods such as Needlet ILC \cite{Delabrouille:2008qd} and its moment \cite{Remazeilles:2020rqw} and Multiclustering \cite{Carones:2022xzs} extensions will be needed. 
Although our quantitative results may be affected,  qualitative conclusions will remain valid as long as the method is still based on ILC.

It would be interesting to relax some of these assumptions and check whether the recommendations presented in section \ref{sec:HWPdesignREC} still ensure that gain calibration for the CMB temperature can mitigate polarization systematics due to the HWP non-idealities. 
We leave this analysis for future work.

\section{Conclusions and perspectives}\label{sec:conclusions_r}
In this work, we presented a simple framework to propagate the HWP non-idealities through the three macro-steps of any CMB experiment: observation of multi-frequency maps, foreground cleaning, and power spectra estimation. 
We focused on the impact of non-idealities on the tensor-to-scalar ratio parameter, $r$.

We generalized the formalism presented in \cite{Monelli:2022pru} to include the polarized Galactic foreground emission (dust and synchrotron), foreground cleaning using a blind method (HILC), bandpass integration, noise, beam smoothing, and gain calibration for the CMB temperature. 
As a concrete working case, we considered a full-sky CMB mission with LiteBIRD-like specifics~\cite{LiteBIRD:2022cnt}.

We validated the code against an ideal HWP and confirmed that the MLE $\hat{r}$ \changedTWO{was compatible with the input value, $r=0.00461$, within the uncertainty}{had negligible bias}. 
Then, we employed more realistic Mueller matrix elements for each of the three telescopes of LiteBIRD and found \changedONE{$\hat{r}=0.0043\pm 0.0005$}{$\hat{r}=(4.30^{+0.56}_{-0.53})\times 10^{-3}$}. 
We showed how the suppression is mostly due to the effective polarization efficiency of the HWP, which averages to a value lower than 1. 
The effective cross-polarization coupling and the foreground residual are found to be subdominant in our output $B$-mode power spectrum.

We found that the bias in $r$ significantly worsens if gain calibration for the CMB temperature is not included in the modeled multi-frequency maps: \changedONE{$\hat{r}=0.0039\pm 0.0005$}{$\hat{r}=(3.94^{+0.52}_{-0.50})\times 10^{-3}$}, which is incompatible with the input value. 
Gain calibration would perfectly remove the HWP effects if $\rho_\text{CMB}^i= g_\text{CMB}^i$ and $\eta_\text{CMB}^i= 0$, which are, however, unrealistic requirements. 
Still, we showed that an effective mitigation can be achieved if we can factorize $\rho_\text{CMB}^i\simeq A^i g_\text{CMB}^i$, we have good knowledge of the $A^i$ coefficients, and $\langle\eta_\text{CMB}^i\rangle\simeq 0$.
These considerations helped us to formulate some recommendations on the HWP design in terms of the HWP Jones parameters:
\begin{itemize}[topsep=2mm,parsep=1mm,itemsep=1mm]
    \item[\small$\triangleright$] Cross-polarization coupling should be small, $\zeta_{1,2}\lesssim 10^{-2}$, which can be achieved for both metal-mesh and multi-layer HWPs;
    \item[\small$\triangleright$]  The loss parameters should also be small, $h_{1,2}\lesssim 10^{-2}$, or, alternatively, $\vert h_1- h_2\vert\lesssim 10^{-3}$;
    \item[\small$\triangleright$]  At least one of $h_{1}(\nu) + h_{2}(\nu)$ and $[1+\cos\beta(\nu)]/2$ should be slowly varying within the band, so that $\rho_\text{CMB}^i\simeq A^i\, g_\text{CMB}^i$;
    \item[\small$\triangleright$]  Cross-polarization coupling can be kept under control by requiring $\zeta_{1,2}$ to be even smaller, or alternatively, by ensuring that $\eta_\text{CMB}^i$ fluctuates around zero.
\end{itemize}
One can characterize the non-idealities of the HWP in laboratory measurements, and a requirement for the smallness of a bias in $r$ gives a requirement for the accuracy of the calibration in the laboratory. 
However, if the above recommendations are implemented in the design of the HWP used for space missions, the in-flight gain calibration for the CMB temperature can also be used to check and correct for the effects of HWP non-idealities in the data, complementing the laboratory calibration.

Some of the recommendations above depend strongly on the class of foreground cleaning methods we used in our end-to-end model. 
We used a blind method (HILC), but if one were to use a parametric component separation method to derive design recommendations, they would likely be different from those listed above. 
This highlights the importance of developing analysis strategies together with hardware designs.

This work represents a first generalization of the model presented in \cite{Monelli:2022pru} towards a more realistic account of how the HWP non-idealities affect the observed CMB. 
However, being semi-analytical, this framework still relies on several simplifying assumptions (see section~\ref{sec:assumptions}). 
One of the most crucial is the isotropy of the foreground SED. 
It would be interesting to relax this assumption and repeat the analysis carried out in this paper, using more elaborate ILC-based methods (e.g., \cite{Remazeilles:2020rqw,Carones:2022xzs}). 
This would help us test the robustness of our recommendations for the design of HWPs in a more realistic context. 
We leave this study for future work.

\acknowledgments
We thank P.\ Campeti, S.\ Giardiello, L.\ Herold, V.\ Muralidhara, M.\ Reinecke, A.\ Ritacco, and Joint Study Group of the LiteBIRD Collaboration for useful discussions.
This work was supported in part by the Excellence Cluster ORIGINS which is funded by the Deutsche Forschungsgemeinschaft (DFG, German Research Foundation) under Germany’s Excellence Strategy: Grant No.~EXC-2094 - 390783311. 
This work has also received funding from the European Union's Horizon 2020 research and innovation programme under the Marie Skłodowska-Curie grant agreement no.\ 101007633. 
TG is supported by World Premier International Research Center Initiative (WPI), MEXT, Japan and by JSPS KAKENHI Grant Number 22K14054. 
The Kavli IPMU is supported by World Premier International Research Center Initiative (WPI), MEXT, Japan.


\appendix

\section{Spectral properties in thermodynamic units}\label{sec:app_a}
For a given frequency $\nu$, anisotropies in specific intensity, $\delta I_\nu$, in units of ${\rm J\,s^{-1}\,m^{-2}\,str^{-1}\,Hz^{-1}}$ and thermodynamic temperature, $\delta T(\nu)$, in units of ${\rm K}$ are related by
\begin{equation}
 \delta I_\nu = \frac{dB_\nu(T_0)}{d T_0}\delta T(\nu)
  = \frac{2\nu^2}{c^2}\frac{x^2e^x}{(e^x-1)^2}k_B\delta T(\nu)\,,
\end{equation}
where $B_\nu(T_0)=2h\nu^3/[c^2(e^x-1)]$ is a black-body spectrum, $x\equiv h\nu/(k_B T_0)$ and $T_0=2.725$ K is the average temperature of the CMB~\cite{Fixsen2009}. 
The thermodynamic temperatures at $\nu$ and at some other reference frequency $\nu_\star$ are related by
\begin{equation}\label{eqn:thermo}
  \delta T(\nu) = \frac{\delta I_\nu}{\delta I_{\nu_*}}\frac{\nu_*^2}{\nu}\frac{x_*^2e^{x_*}}{(e^{x_*}-1)^2}\frac{(e^x-1)^2}{x^2e^x}\delta T(\nu_*)\,.
\end{equation}
The specific intensity of CMB anisotropies follows a differential black-body, while dust and synchrotron can be modeled as a modified black-body and a power law, respectively \cite{Planck:2018yye}
\begin{subequations}\label{eqn:dustandsynch}
\begin{align}
    \delta I_{\text{CMB},\nu} &= \frac{2\nu^2}{c^2}\frac{x^2e^x}{(e^x-1)^2}k_B\,\delta T\,,\\
    \delta I_{\text{dust},\nu} &= A_\text{dust} \left(\frac{\nu}{\nu_\filledstar}\right)^{\beta_\text{dust}} B_\nu(T_\text{dust})\,,\\
    \delta I_{\text{sync},\nu} &= A_\text{sync} \left(\frac{\nu}{\nu_\smallstar}\right)^{\beta_\text{sync}}\,.
\end{align}
\end{subequations}
By plugging these expressions in eq.\ \eqref{eqn:thermo}, we obtain the SED of CMB, dust, and synchrotron in terms of the CMB thermodynamic temperature:
\begin{subequations}
\begin{align}
    \delta T_{\text{CMB}}(\nu) &= \delta T_\text{CMB}\,,\\
    \delta T_{\text{dust}}(\nu) &= \left(\frac{\nu}{\nu_\filledstar}\right)^{\beta_\text{dust}} \frac{B_\nu(T_\text{dust})}{B_{\nu_\filledstar}(T_\text{dust})} \frac{\nu_\filledstar^2}{\nu^2}\frac{x_\filledstar^2e^{x_\filledstar}}{x^2e^{x}} \frac{(e^{x}-1)^2}{(e^{x_\filledstar}-1)^2}\,\delta T_\text{dust}(\nu_\filledstar)\,,\\
    \delta T_{\text{sync}}(\nu) &= \left(\frac{\nu}{\nu_\smallstar}\right)^{\beta_\text{sync}} \frac{\nu_\smallstar^2}{\nu^2}\frac{x_\smallstar^2e^{x_\smallstar}}{x^2e^{x}} \frac{(e^{x}-1)^2}{(e^{x_\smallstar}-1)^2}\,\delta T_\text{sync}(\nu_\smallstar)\,.
\end{align}
\end{subequations}

\section{Relating Mueller to Jones parameters}\label{sec:muellerjones}
Mueller and Jones calculus are two different matrix methods to describe and manipulate polarized radiation. 
Mueller calculus works with intensities, while Jones calculus works directly with the $x$ and $y$ components of the electric field. 
Any Jones matrix, $J$, can be transformed into the corresponding Mueller–Jones matrix $\mathcal{M} = A\,(J \otimes J^*)\,A^{-1}$, where
\begin{equation}
\mathrm{A} =
  \begin{pmatrix}
    \phantom{-}1\phantom{-} &  \phantom{-}0\phantom{-} & \phantom{-}0\phantom{-} &  \phantom{-}1\phantom{-} \\
    1 &  0 & 0 & -1 \\
    0 &  1 & 1 &  0 \\
    0 &  i & -i &  0 \\
\end{pmatrix}\,.
\end{equation}
Here, $*$ denotes the complex conjugate and $\otimes$ is the Kronecker product.
The Jones matrix for a non-ideal HWP is
\begin{equation}
    J_\textsc{hwp} = \begin{pmatrix}
        1 + h_1 & \zeta_1 e^{i\chi_1} \\
        \zeta_2 e^{i\chi_2} & -(1 + h_2) e^{i\beta}
    \end{pmatrix}\,,
\end{equation}
where $h_{1,2}$ are loss parameters describing the deviation from the unitary transmission of $E_{x,y}$; $\beta$ parametrizes the deviation from $\pi$ of the phase shift between $E_x$ and $E_y$; $\zeta_{1,2}$ and $\xi_{1,2}$ describe the amplitude and phase of the cross-polarization coupling. 
All Jones parameters tend to zero in the ideal limit.

\bibliographystyle{JHEP}
\bibliography{main}

\end{document}